# Blindfolded monkeys or financial analysts: who is worth your money?[1]

Giuseppe Pernagallo† and Benedetto Torrisi‡

*Collegio Carlo Alberto, Piazza Vincenzo Arbarello 8, 10122 Turin, Italy*
†*email: giuseppepernagallo@yahoo.it (correspondent author)*
*University of Catania, Department of Economics and Business, 55, 95129, Catania*
‡*email: btorrisi@unict.it*

**Abstract**

The efficient market hypothesis has been considered one of the most controversial arguments in finance, with the academia divided between who claims the impossibility of beating the market and who believes that it is possible to gain over the average profits. If the hypothesis holds, it means, as suggested by Burton Malkiel, that a blindfolded monkey selecting stocks by throwing darts at a newspaper's financial pages could perform as well as a financial analyst, or even better. In this paper we use a novel approach, based on confidence intervals for proportions, to assess the degree of inefficiency in the S&P 500 Index components concluding that several stocks are inefficient: we estimated the proportion of inefficient stocks in the index to be between the 12.13% and the 27.87%. This supports other studies proving that a financial analyst, probably, is a better investor than a blindfolded monkey.

**Keywords:** Confidence interval; Efficient market hypothesis; Informational efficiency; Random walk; Runs test; Variance ratio.

## 1. Introduction

The efficient market hypothesis has been considered one of the most controversial arguments in finance, with the academia divided between who claims the impossibility of beating the market and who believes that it is possible to gain over the average profits. The amount of empirical studies produced on the issue is extraordinary and trying to assess the problem with originality is a difficult task. However, in this paper we propose a new approach to test the degree of inefficiency of the stock

---

market providing new interesting results supporting the existence of inefficiencies.

The random walk model, which is probably the most versatile model for empirical validation of the EMH, states that in a security's price series, all subsequent price changes are random departures from past prices (Malkiel, 2003). If information is immediately reflected in stock prices, therefore information today cannot be used to predict prices tomorrow and the path of price changes will be completely random and unpredictable. Using a vivid image suggested by Burton Malkiel (1973) "*it means that a blindfolded monkey throwing darts at a newspaper's financial pages could select a portfolio that would do just as well as one carefully selected by the experts*". The irreverent image of Burton Malkiel seems a joke but was taken seriously by the San Francisco Chronicle. This curious anecdote (Mishkin and Eakins, 2012, p. 127) shows perfectly the meaning of the efficient market model: it was asked to eight analysts to pick five stocks at the beginning of the year and then their results were compared with those obtained by the stocks chosen by Jolyn, an orangutan. Surprisingly (or maybe not), Jolyn beat the analysts as often as they beat her. This is not, of course, a scientific way to test the model, but such absurdities are not infrequent in financial markets.

We should warn the reader that the approach followed in this study is intentionally very conservative, since the supporters of stock markets efficiency have provided strong proofs of their claims. Nevertheless, our strong and prudent statistical framework consents us to affirm that some degree of inefficiency is still present, at least in the U.S. stock markets. We conclude therefore that for many assets the weak form of the efficient market hypothesis is rejected.

The paper is organized as follows. Section 2 defines the notion of market efficiency and presents the main empirical literature. Section 3 presents the adopted methodology and results. The final section concludes.

## 2. A brief review of previous studies on the Efficient Market Hypothesis

Informational efficiency is the situation that characterizes a market where prices fully reflect all available information. The Nasdaq glossary defines informational efficiency as: *"The degree to which market prices correctly and quickly reflect information and thus the true value of an underlying*

*asset.²"*. The most diffused expression to indicate this condition is "efficient market hypothesis" (EMH). This kind of definition is, in most of cases, split into three different degrees of efficiency: weak, semi-strong and strong form. This categorization is useful because consents us to test the degree of informational efficiency from the lowest one to the highest one. The classic definitions of the three forms are:

1. weak form: past prices are already incorporated into current prices; therefore, they cannot be used to gain excess returns (if this form holds, technical analysis is completely useless);
2. semi-strong form: security prices adjust to publicly available information (such as annual reports, new security issues etc.) very rapidly to reflect also these information (if this form holds, fundamental analysis becomes almost useless);
3. strong form: a market where prices reflect all kind of information, public and private, so that is impossible to earn excess returns.

The strong form incorporates the other two whereas the semi-strong form incorporates the weak form. One way to test the degree of informational efficiency is to test whether stock market analysis is capable to ensure above the average profits. The two basic approaches to invest in stock markets are technical analysis and fundamental analysis.

Technical analysis is based essentially on the study of stocks' charts, for this reason the supporters are called *chartists*. This methodology is grounded in the "castle-in-the-air theory" (Malkiel, 1973): it is possible to understand better market dynamics using psychological factors instead than the "intrinsic value" of a firm, studying the behaviour of investors and trying to predict it. The basic idea is that in good times investors tend to be optimistic (they build castles in the air) whereas in bad times they are pessimistic; this is reflected by the path of a stock price, so the analysis of trends can reveal precious information over its future evolution. Edwards and Magee (1948) said that "*Prices move in trends, and trends tend to continue until something happens to change the supply-demand balance*";

---
² *https://www.nasdaq.com/investing/glossary/i/informational-efficiency*

for chartists it is possible to use past information to make successful predictions. The intuition behind the idea of Edwards and Magee was in some sort of sense comparable with the one proposed by Mandelbrot (1963, p. 418), who observed that "*large changes tend to be followed by large changes—of either sign—and small changes tend to be followed by small changes*", a phenomenon known as volatility clustering. In truth, this approach may have a sense especially in the case of self-fulfilling prophecies, a direct consequence of herd behaviour. Suppose that a sentiment or a news, even completely unjustified, spreads in the market, if enough traders will believe it, they will cause with their actions the prevented result. A trader able to anticipate these situations can outperform the market.

The second paradigm is called fundamental analysis because the analyst tries to evaluate the real value of a stock, called intrinsic value, and then compare this value with the market value to understand if it is overvalued, undervalued or correctly valued by the market. This methodology is grounded in the "firm-foundation theory" (Malkiel, 1973), i.e. it uses information relative to the firm (e.g. dividends payments, annual or infra-annual reports, strategic decision announcements etc.) to estimate the evolution of its cash flows over a given period and then actualize them to correctly value the stock. If, for instance, the stock is undervalued the analyst will buy it before the market will adjust its price, so that when the price will go up, she can sell the stock and gain the differential.

It is clear now why the weak form of EMH is destructive for technical analysis whereas the semi-strong for the fundamental analysis: in the first case, information from past prices is useless to outperform the market since it is already reflected in present prices; in the second case, if prices adapt rapidly to public information, the computation of a fundamental value become a worthless exercise since the market is already evaluating the stock price correctly. Fama (1965a) shows how the random walk model presents important challenges to the followers of both technical analysis and fundamental analysis.

Some evidences seem to support the EMH, for example, it has been showed in several works (Jensen, 1968; Malkiel, 1995) that mutual funds have not been able to outperform the market. Other studies (Ball and Brown, 1968; Fama et al., 1969) showed that, coherently with the semi-strong form, positive announcements about a company do not produce, on average, an increase in the price of its stock because this information is already incorporated in the stock price. Fama et al. (1969) concluded that the stock market is efficient with respect to its ability to adjust to the information implicit in a stock split using the method of residuals analysis. Scholes (1969) used the same method on a sample of 696 new issues of common stock during the period 1926-1966 to provide further evidence in favour of the semi-strong form efficiency. Several works have tested the overall performance of technical analysis showing that it is not able to outperform the market (Alexander, 1961; Allen F. and Karjalainen, 1999).

Evidences against the EMH have gained popularity in recent years, among these we have the *small firm effect*, that occurs when small firms can earn abnormal returns over long periods of time (Reinganum, 1983). Various explanations have been made referring to tax issues, large information costs related to the valuation of small firms or low liquidity of their stocks, but the existence of the phenomenon is in contradiction with the EMH.

One of the better-known anomalies in stock markets is the so-called *January effect*, i.e. the fact that stock prices have tended to show a consistent price increase from December to January, making this phenomenon predictable and hence contradicting the random walk hypothesis. The amount of studies concerned with this anomaly is remarkable (Reinganum, 1983; Chan, 1986; Thaler, 1987; Eakins and Sewell, 1993) and many authors explain the phenomenon principally using tax considerations. Investors tend to sell stocks in December because they can then take capital losses on their tax return and reduce their tax liability. In January, they can repurchase the stocks, raising up their prices and producing higher returns. This theory seems reasonable but does not explain why institutional investors, that are income tax-free, do not take advantage of the higher returns in January buying stocks in December, causing an increase in those prices and absorbing the anomalous returns.

Other anomalies that seem to contradict the EMH are the stickiness in pricing errors correction after news announcements (market overreaction; see, for example, De Bondt and Thaler, 1987 or

Chopra et al., 1992) and the related phenomenon of fluctuations in stock prices much greater than fluctuations in their fundamental value (*excess volatility*; see, for example, Shiller, 1981). Among the others, LeRoy and Porter (1981) pointed out that stock markets show excess volatility and they rejected market efficiency.

Even if it is controversial, mean reversion of stock returns is another feature of financial markets which falsifies the assumptions of the EMH. Summarily, stock returns exhibit the tendency to return to their means through time thus making their movements partly predictable in contrast to the random walk model of stock prices (for a summary, see Engel and Morris, 1991). As noted by Peron and Vodounou (2005), a common conclusion is that the EMH is rejected when using long horizon returns (generally 3 to 10 years), which implies a mean-reverting behaviour for prices.

Keim and Stambaugh (1986) found statistically significant predictability in stock prices by using forecasts based on certain predetermined variables. Lo and MacKinlay (1988) rejected the random walk hypothesis for weekly stock market returns using the variance ratio test, furthermore, also Jegadeesh (1990) provided strong evidence in favour of predictability of securities returns.

The strong form efficiency seems to be more theoretical than realistic. Niederhoffer and Osborne (1966) found that specialists on the N.Y.S.E. (from now on, NYSE) apparently use their monopolistic access to information. Scholes (1969) evidenced that officers may have monopolistic access to information about their corporations and even Fama (1970) warned about the unrealistic feasibility of this extreme hypothesis; furthermore, real cases seems to falsify the EMH in its strong form (for example, the case of Ivan Boesky and the insider trading scandal which involved him).

A great challenge for the EMH is represented by the existence of speculative bubbles: with this term we indicate a situation of a great discrepancy between the true value of a stock and its current market value. If markets are informationally efficient, stock price should always reflect the real value of that security because of the rational expectation assumption. This phenomenon can assume catastrophic proportions, as happened with the Dot-com Bubble, when the euphoric hunger of investors pushed up the stock prices of internet-based company far beyond any realistic considerations. In this case the EMH fails to provide a better explanation than behavioural finance: cognitive bias and behavioural parameters seem to be the main drivers of such phenomena.

## 3. Measuring the level of informational efficiency of the U.S. stock market

### 3.1 Data and methodology

Our study focuses on the U.S. stock market, because of its relevance, and consists of several tests to measure the level of informational efficiency of the traded stocks. Measuring the level of informational efficiency is possible assessing the randomness of returns. Given the massive amount of stocks traded in the U.S. stock market, we choice to study only stocks that are part of the S&P 500 Index.

The index was created in 1923 by Standard and Poor's and covered at that time 233 companies[3]. The S&P 500 Index in its composition of 500 companies was introduced in 1957 and it is computed as it follows

$$\text{Index level} = \frac{\sum_i P_i * Q_i}{\text{Divisor}} \quad (1)$$

where the numerator is the sum of the price of each stock in the index multiplied by the number of shares used in the index calculation, and the denominator is the "divisor", a quantity used to consider particular events, such as spin-offs or structural changes, to avoid the alteration of the numerical value of the index[4]. This formula descents from a modified Laspeyres index and is a "*base-weighted aggregative*" method (Cowles, 1939) with the consequence that the index is weighted toward companies with large market capitalization. The number of the included companies and their importance for the U.S. market make the index a good indicator of the performance of the American stock market consenting us to have a good overview of the overall level of efficiency.

The survey covers the period from January 2, 2008 (first trading day) to July 24, 2018 and data are gathered on daily basis (trading days). However, we considered only the S&P 500 Index components that were continuously part of the index for the entire period, which is essential to generalize the results for all the stocks.

---

[3] https://www.reuters.com/article/us-usa-stocks-sp-timeline/timeline-key-dates-and-milestones-in-the-sp-500s-history-idUSBRE92R11Z20130328
[4] Index Mathematics Methodology, available at https://us.spindices.com/documents/methodologies/methodology-index-math.pdf.

The S&P 500 Index consists of 505 components or stocks; using *Thomson Reuters Eikon & Datastream* and *Wikipedia* (which refers to *Compustat database* and the *S&P website*) we retraced the history of the S&P 500 from September 1989 (the oldest year available on *Datastream*) to July 2018. Therefore, we eliminated from our analysis the components which were added, eliminated or substituted within the period, which leads us from 505 components to 305 components continuously traded from 2008 to 2018 as part of the S&P 500 Index. These 305 components were our target population.

We found out that, of the 305 components, the 11.48% are original stocks (35 stocks), in the sense that they were part of the original composition of the index in 1957[5]. We obtain this information using the Appendix in Siegel's book (2005), which reports the original S&P 500 Index firms. We considered only those firms that did not experience mergers, acquisitions or drastic changes in their original structure. Each stock is classified per GICS[6] (Global Industry Classification Standard) taxonomy; as shown in Table 1, the composition of the index is dominated by financial corporations (17,70%), whereas telecommunication services and real estate corporations are the least represented (0.98% and 3.93% respectively). Interestingly, of the 35 survivor companies, no one possesses financial services as core activity, whereas from 2008-2018 most of the stocks are issued by financial companies with a growing component of companies involved in the information technology services.

---

[5] We know that these 35 stocks have been continuously part of the index from September 1989 to July 2018, except for Macy's. We were not able to reconstruct all their history because of the lack of data.
[6] They are classification standards published by MSCI in collaboration with Standard & Poor's to facilitate a globally homogenous criterion of classification. Companies are classified on the basis of their core business and inserted in one of the 11 provided categories.

**Table 1.** Distribution of the S&P 500 Index components by GICS for the period 2008-2018 (from January 2, 2008 to July 24, 2018).

|    | GICS                       | Absolute Frequency | Relative frequency (%) |
|----|----------------------------|--------------------|------------------------|
| 1  | Consumer Discretionary     | 45                 | 14.75%                 |
| 2  | Consumer Staples           | 26                 | 8.52%                  |
| 3  | Energy                     | 18                 | 5.90%                  |
| 4  | Financials                 | 54                 | 17.70%                 |
| 5  | Health Care                | 34                 | 11.15%                 |
| 6  | Industrials                | 38                 | 12.46%                 |
| 7  | Information Technology     | 38                 | 12.46%                 |
| 8  | Materials                  | 15                 | 4.92%                  |
| 9  | Real Estate                | 12                 | 3.93%                  |
| 10 | Telecommunication Services | 3                  | 0.98%                  |
| 11 | Utilities                  | 22                 | 7.21%                  |
|    | Tot                        | 305                | 100%                   |

Source: our elaboration based on *Thomson Reuters Eikon & Datastream data*.

241 (79.02%) stocks of the 305 under analysis are traded on the NYSE (New York Stock Exchange), whereas the remaining 64 (20.98%) are traded on the NASDAQ.

Given the impossibility of repeating all the tests for each of the 305 stocks, we extracted a sample from the 305 stocks, performed the desired tests and concluded if that stock is informationally efficient or not. Once we have repeated the procedure for all the sampled stocks, we obtained a proportion of the inefficient stocks in the S&P 500 Index and tried to generalize the results to the entire index via confidence intervals.

### 3.2 The sampling scheme

A crucial step is the determination of the sample size. We must note that the presented conclusions hold only for the stocks in the S&P 500 Index; as noted by Fama (1965b), the component companies of the main indexes are among the largest and most important in their fields, therefore they are not a random sample of stocks from the NYSE and the NASDAQ, so the empirical results will be strictly applicable only to the shares of large important companies. In his work, Fama selected all the stocks of the Dow-Jones Industrial Average (DJIA) at the time, but the choice seems to be based only on the convenience of dealing with a small set of stocks. Kendall (1953) used 22 price-series which were selected for the purpose of his study, without specifying how that number was determined. Jensen (1968) studied the performance of 115 open end mutual funds; the sample size chosen by the author is probably related to data availability.

Following Cochrane (1953), if the units are classified into two classes, $C$ and $C'$ (informationally efficient stocks and non-efficient stocks), given $N$ large enough, a first approximation of the sample size is

$$n_0 = \frac{t^2 pq}{d^2} \quad (2)$$

where $t$ is the abscissa of the normal curve which cuts off an area $\alpha$ at the tails (usually indicated in textbooks as $z_\alpha$), $d$ is the chosen degree of precision or margin of error in the estimated proportion $p$ of units in class $C$, and $q$ is the complementary of $p$, or $(1-p)$. The ratio between $t^2$ and $d^2$ is the desired variance of the sample proportion. If $n_0/N$ is negligible, then $n_0$ is a good approximation of the sample size; if not, the following formula can be used

$$n = \frac{n_0}{1 + \frac{n_0}{N}} \quad (3)$$

A simpler formula and more suitable for our purposes can be derived from (2). The sample size formula adopted is (4) and is generally known as Yamane formula (1967a; 1967b)

$$n = \frac{N}{1 + Nd^2} \quad (4)$$

and its derivation is shown in Appendix A (Supplementary Material online) of the present work. The use of this formula is based on the following assumptions: the sample percentage $p$ is assumed to be normally distributed, a confidence of 95% is assumed and $p$ is set equal to 0.5.

The specification of the desired precision is a controversial task, Cochrane (1953) admitted that a certain level of arbitrary is involved in the survey design, and the final decision relies mostly on the purpose of the study. We think that a reasonable value is 0.1, which avoids a too large or too small sample size. Finally, the chosen sample size is approximately 75, which means that we will select 75 stocks from the S&P 500 Index components in the period 2008-2018.

We stratified our sample to consider effects on the stocks' performance due to the sector and the exchange market. The distinctions between stocks traded on the NYSE and the NASDAQ are very important. First, the NYSE is older and bigger in terms of market capitalization of the listed firms than the NASDAQ. Second, NYSE is an auction market where brokers purchase stocks on behalf of firms or clients. Trades are concluded between individuals on the floor of the exchange, whereas NADSDAQ is a computer-based market with orders processed electronically; this distinction can be important in our analysis concerning information efficiency. Finally, companies on the NYSE are perceived to be less volatile because its composition includes principally blue-chip companies, considered to be more stable and established.

There are two layers of stratification, the first one regarding the stock exchange (NYSE or NASDAQ) and the second one the sector of the stock. To consider a more parsimonious distribution of the stocks in the second stratus and avoid empty strata, we merge the 11 GICS categories into 5 categories based on similarities. Table 2 synthesizes the second layer of stratification. The subsample size from each stratum is proportional to the size of the subpopulation in the stratum (*proportional allocation*); the sampling scheme is represented in Figure 1. For example, knowing that the Health Care sector weights the 11.62% of the stocks traded on the NYSE, we have that the subsample must be 59*0.1162 ≈ 7 (rounded to the nearest integer) to maintain the proportion and the representativeness in the sample. Using the same rationale for all the subsamples, we selected, randomly, 14 stocks traded on the NYSE issued by Consumer Discretionary and Staples companies, 3 stocks traded on the NASDAQ issued by Consumer Discretionary and Staples companies and so on.

**Table 2.** Distribution of the S&P 500 Index components by stock exchange and merged GICS categories for the period 2008-2018 (weights in brackets).

|   | GICS | Traded on NYSE | Traded on Nasdaq | Tot |
|---|---|---|---|---|
| 1 | Consumer Discretionary and Staples | 57 (23.65%) | 14 (21.88%) | 71 |
| 2 | Energy, Industrials, Materials and Utilities | 87 (36.10%) | 6 (9.38%) | 93 |
| 3 | Financials and Real Estate | 57 (23.65%) | 9 (14.06%) | 66 |
| 4 | Health Care | 28 (11.62%) | 6 (9.38%) | 34 |
| 5 | Information Technology and Telecommunication Services | 12 (4.98%) | 29 (45.31%) | 41 |
|   | Tot | 241 | 64 | 305 |

Source: our elaboration on *Thomson Reuters Eikon & Datastream data*.

**Figure 1.** The sampling scheme of the study using two strata: stock exchange (NYSE or NASDAQ) and GICS (five categories as showed in Table 2).

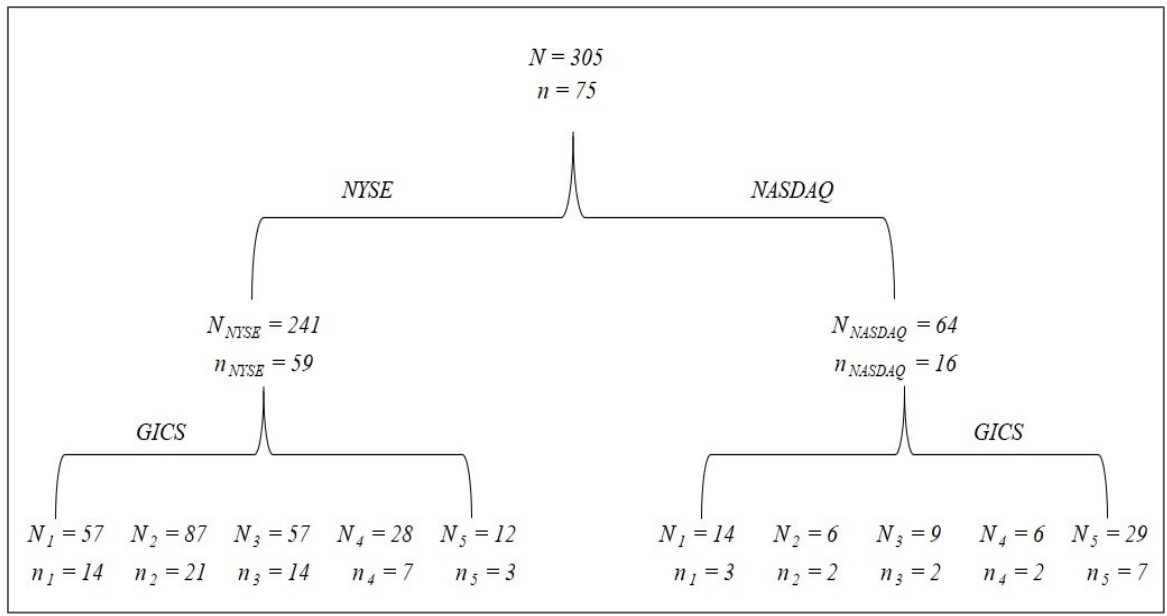

*Source: our elaboration.*

This stratification is just one of the possible, but it results resonable. For example, stratifying for market capitalization is not appropriate becuase several companies in a period of more than 10 years experienced a drammatic change in their capitalization; stratifying for original companies and new companies is not appropriate beacuse stocks can be removed and inserted several times in the index, besides, the low number of original companies would create empty strata.

### 3.3 Definition of financial returns

We gathered the time series of close prices of the selected stocks adjusted for all applicable splits and dividend distributions as proposed in Fama (1965b) using *Yahoo! Finance* data[7]. The period covered goes from December 31, 2007 to July 24, 2018 (included) for a total of 2660 observations. The tests are not performed on daily prices but on the first differences of their natural logarithms. We define this variable "*financial return*" or "*log-return*"

$$r_{t+1} = \log_e p_{t+1} - \log_e p_t \qquad (5)$$

where $p_{t+1}$ is the price of the security at the end of day $t + 1$, and $p_t$ is the price at the end of day $t$. This cause the loss of an observation; hence the log-returns series covers the period from January 2, 2008 to July 24, 2018 (2659 observations).

### 3.4 Runs test

Runs test is a common methodology used to test the randomness in price changes (see, for example, Fama 1965b). A run can be defined as an unbroken sequence of similar events or like objects (Bradley, 1968). For example, in the series AABABBBAA there are five runs: one run of A's of length 1, two runs of A's of length 2, one run of B's of length 1 and one run of B's of length 3. If the outputs A and B are randomly distributed, we should not recognize any pattern in the series. The hypotheses are

$$H_0 : \text{data are randomly distributed}$$
$$H_1 : \text{data are not randomly distributed.}$$

The test statistic is

$$Z = \frac{\textit{Observed no. runs - Expected no. runs}}{\textit{SD no. runs}}$$

where, given $m$ events of the type 1 and $n$ events of the type 2, and $m + n = N$

$$\textit{Expected no. runs} = \frac{2nm}{N} + 1$$

and

---

[7] Details of these adjustments are available at https://help.yahoo.com/kb/SLN28256.html.

$$SD \text{ no. runs} = \sqrt{\frac{2mn(2mn - N)}{N^2(N - 1)}}$$

with SD the standard deviation of the number of runs. We reject the null hypothesis of randomness if the computed statistic, in modulus, is greater than the critical value (1.96 with the classical alpha of 5%) or if *p-value < significance level*. We used this nonparametric test in two different ways. First, we wanted to test if it is not possible to obtain above average returns which implies informational efficiency for the analysed stock. We define

$$d_t = \begin{cases} 1, & r_t > \mu \\ 0, & r_t \leq \mu \end{cases}$$

as the dichotomized series of log-returns. In this way we have obtained a sequence of runs (with symbols 1 and 0) and we can apply the test of randomness; if we reject the null hypothesis of randomness, we can conclude that above average returns are not randomly distributed and hence they can be predicted. Second, to compute the runs up and down for $N$ numeric data values (log-returns), the sign of the difference between each value and the previous one is recorded and used to create a binary series of $N-1$ signs, indicating with + a sequence of increasing returns and with – a sequence of decreasing returns. If we reject the null hypothesis, we can conclude that increases and decreases in the financial returns series are not random and hence they can be predicted.

**3.5 Variance ratio test**

The Lo and MacKinlay variance ratio test (1988) uses the fact that the variance of the increments of a random walk is linear in the observation interval. That is, given the random walk $X_t$, the variance of $X_t - X_{t-2}$ is twice the variance of $X_t - X_{t-1}$. Hence, if the natural logarithm of the stock price follows the random walk version of the martingale hypothesis, then the return variance should be proportional to the return horizon (Fong, Koh and Ouliaris, 1997; Chang and Ting, 2000). If we define $X_t = \ln P_t$, with $P_t$ the price of a stock, under the martingale null hypothesis, the variance of the $q$-th holding-period returns, $X_t - X_{t-q}$, should be $q$ times the variance of the one-period returns, $X_t - X_{t-1}$. The null hypothesis therefore implies that the variance ratio

$$V(q) = \frac{\sigma^2(q)}{\sigma^2} \qquad (6)$$

is equal to 1, because the numerator is $1/q$ times the variance of the $q$-th period returns and the denominator is the variance of one-period returns (Fong, Koh and Ouliaris, 1997). Lo and MacKinlay (1988) derived a version of the specification test of the random walk model that is robust to changing variances (heteroskedasticity) given the growing consensus among financial economists, that volatilities of asset returns do change over time (Merton, 1980; Poterba and Summers, 1986; Bollerslev, Chou, and Kroner, 1992). This version of their test is robust to general forms of conditional heteroscedasticity, including *ARCH* processes and deterministic changes in the variance. The Lo and MacKinlay's heteroscedasticity-robust standard normal test-statistics is indicated with $z^*(q)$; Lo and MacKinlay's (1989) show that the asymptotic distribution of $z^*(q)$ performs well in finite samples, and the variance ratio test is more reliable than Dickey-Fuller and Box-Pierce tests. With this kind of tests, it is possible to uncover low-frequency variations in the data such as the presence of long memory or mean reversion (Fong, Koh and Ouliaris, 1997).

As noted by Fong, Koh and Ouliaris (1997), because the researcher has to examine variance ratios across all preselected lags, the null hypothesis is not rejected only if none of the variance ratio estimates are significantly different from 1. Since it is possible that some of the variance ratios estimates across different lags differ from 1, it is appropriate not only to focus on the significance of individual tests, but also control for the joint test size. To control the size of the joint test, Chow and Denning (1993) proposed a (conservative) test statistic that examines the maximum absolute value of a set of multiple variance ratio statistics. The p-value for the Chow-Denning statistic is bounded from above by the probability for the Studentized Maximum Modulus (SMM) distribution with parameter $m$ and infinite degrees-of-freedom in order to approximate this bound with the asymptotic SMM distribution.

## 3.7 Data analysis and results

The tables in this section show the name of each randomly selected stock preceded by a label for each subsample, where NYSE and NASD are the abbreviation for the stock exchange and the numbers refer to the aggregated GICS category; so, for example, NASD2 indicates a stock traded on the NASDAQ issued by a company operating in the energy, industrials, materials or utilities sector.

To detect if above average returns and changes in the sign of returns behave as a random series we tested if the runs above and below the mean and the series of runs up and down are randomly distributed. This is an easy way to measure the level of informational efficiency of each stock since the more random data are, the more efficient and unpredictable the series. Table 3 shows the results for the two variations of the test. Asymptotic p-values are computed for runs above and below the mean, whereas for runs up and runs down the length of runs and what sequence of runs is significant (based on p-value) are reported. "Runs up" refers to a sequence of increasing returns such as -0.2, -0.1, 0, 0.1, 0.2; "runs down" refers to a sequence of decreasing returns such as 0.2, 0.1, 0, -0.1, -0.2. If no change occurs at start of the series, the run is counted as an up; if no change occurs after an up, the runs is counted as an up; finally, if no change occurs after a down, the run is counted as a down. The length of runs up or down is the number of symbols in succession without interruptions. For example, for Best Buy the length of runs up goes from 1 to 5, which means that the sequence of sign recorded from the series are +, + +, + + +, + + + +, + + + + +, and with the 3** we indicate that the sequence of signs made by three ups is significantly (at 5%) non-random.

Even if the interpretation of the test for runs above and below the mean is straightforward, the interpretation of runs up and down is difficult: how many runs should be significant to consider the series non-random? We simulated 1,000 random series and performed the test to observe whether the results were similar to these outputs. For all the simulated random series the test of runs above and below the mean was almost always able to catch the random nature of data and only in few occasions it resulted significant: 6 times the test signalled, wrongly, significance at 1%, and in two of these cases also some runs up or down resulted significant (but no more than three sequences). The test of runs up and down failed with a higher frequency and in some simulations even 4 sequences were significant. We can therefore conclude that the test of runs above and below the mean is particularly reliable, especially when the significance is at 1%, whereas the test of runs up and down must be interpreted with prudence and conclude for the rejection of the null hypothesis if also the other test suggests the same or if there are more than 3 sequences of runs highly significant.

The evidence against non-randomness from our data is remarkable. The test over runs above and below the mean shows strong evidence against randomness in the series of TJX, Bank of New York Mellon, Loews, U.S. Bancorp, Huntington Bancshares and Fiserv. Interesting results concerning the analysis of runs up and down characterize the series of Occidental Petroleum, Southwest Airlines, Waste Management and Franklin Resources, where numerous runs sequences are highly significant indicating non-randomness. The results of the two versions of the test are in some cases concordant (for example, Fiserv and Southwest Airlines), in other cases they reach different conclusions. It is the case of Cintas (there is no evidence of randomness if we use runs up and runs down, whereas strong evidence results from the analysis of runs above and below the mean) or Waste Management (the opposite case). However, as pointed out by Fama (1965b), the runs tests are rigid in their approach in determining the duration of upward and downward movements in prices; indeed, a run is terminated whenever there is a change in sign in the sequence of price changes, regardless of the size of the price change that causes the change in sign. The test over runs above and below the mean presents also two limits: it is dependent on the temporal interval under analysis since the mean is computed over the whole interval, with the consequence that shorter or larger interval of time could provide completely different conclusions; the mean of financial returns can also be negative, indicating a loss. When the mean is negative, from the point of view of an investor, it is of little interest to know that it is possible to obtain above average returns, because also irrelevant returns are higher than a loss. Fortunately, the use of stocks included in an important index such as the S&P 500 reduces the cases of negative means. Both the tests considering the S&P 500 Index returns series suggest non-randomness.

**Table 3.** Runs test above and below the mean for the dichotomized series of returns (symbols: 1 and 0) and runs up and down for changes in the sign of subsequent returns (symbols: + and −); ns stands for "not significant". The results of the tests are obtained via the spreadsheet provided by Dr. Martin Sewell.

|  | Stock | Runs above and below the mean Asymptotic p-value | Runs up Lenght | Runs up Significant | Runs down Lenght | Runs down Significant |
|---|---|---|---|---|---|---|
| NYSE1 | Best Buy | 0.9363 | 1-5 | 3** | 1-5 | ns |
| NYSE1 | Campbell Soup | 0.0265** | 1-5 | 4* | 1-5 | ns |
| NYSE1 | Carnival | 0.1098 | 1-6 | ns | 1-6 | 6*** |
| NYSE1 | Clorox | 0.0217** | 1-5 | ns | 1-5 | 2** |
| NYSE1 | Constellation Brands A | 0.8578 | 1-5 | 4* | 1-2-3-4-5-7 | 5**-7*** |
| NYSE1 | General Mills | 0.0163** | 1-5 | 3** | 1-6 | ns |
| NYSE1 | Kroger | 0.2349 | 1-5 | ns | 1-7 | 7*** |
| NYSE1 | Newell Brands | 0.1265 | 1-5 | ns | 1-6 | ns |
| NYSE1 | Nordstrom | 0.4660 | 1-4 | 2*-5* | 1-5 | ns |
| NYSE1 | Sysco | 0.0820* | 1-5 | ns | 1-5 | 4** |
| NYSE1 | TJX | < 0.0001*** | 1-4 | 5* | 1-5 | ns |
| NYSE1 | Walmart | 0.0152** | 1-5 | ns | 1-5 | ns |
| NYSE1 | Walt Disney | 0.4259 | 1-5 | 2** | 1-5 | ns |
| NYSE1 | Yum! Brands | 0.0267** | 1-5 | 2**-4** | 1-5 | ns |
| NYSE2 | MMM | 0.0688* | 1-6 | 6** | 1-5 | ns |
| NYSE2 | AES | 0.0472** | 1-6 | 1*-3* | 1-6 | ns |
| NYSE2 | American Electric Power | 0.0741* | 1-5 | ns | 1-5 | ns |
| NYSE2 | Cummins | 0.1932 | 1-6 | ns | 1-2-3-4-5-7 | 4*-7*** |
| NYSE2 | Eaton | 0.8919 | 1-6 | 6** | 1-5 | 3*** |
| NYSE2 | Exelon | 0.0729* | 1-5 | ns | 1-5 | ns |
| NYSE2 | Exxon Mobil | 0.0055*** | 1-6 | 3*-4* | 1-5 | ns |
| NYSE2 | FirstEnergy | 0.4486 | 1-2-3-4-5-7 | 7*** | 1-5 | ns |
| NYSE2 | General Dynamics | 0.0164** | 1-5 | ns | 1-6 | ns |
| NYSE2 | Hess | 0.9226 | 1-6 | ns | 1-5 | ns |
| NYSE2 | L3 Technologies | 0.9664 | 1-5 | ns | 1-6 | 2* |
| NYSE2 | Nisource | 0.4388 | 1-2-3-4-5-8 | 3**-8*** | 1-5 | 2* |
| NYSE2 | Noble Energy | 0.0345** | 1-6 | ns | 1-5 | 2** |
| NYSE2 | Occidental Petroleum | 0.2645 | 1-6 | 4*-5*** | 1-5 | 2**-4* |
| NYSE2 | Parker-Hannifin | 0.3203 | 1-5 | ns | 1-6 | 3** |
| NYSE2 | Pinnacle West Capital | 0.7153 | 1-6 | ns | 1-5 | ns |
| NYSE2 | Schlumberger | 0.4968 | 1-5 | ns | 1-6 | ns |
| NYSE2 | Southwest Airlines | 0.0036*** | 1-5 | 1**-3** | 1-5 | 1**-2*-5** |
| NYSE2 | Textron | 0.0509* | 1-5 | 1**-2** | 1-5 | 5*** |
| NYSE2 | Vulcan Materials | 0.8005 | 1-5 | ns | 1-6 | ns |
| NYSE2 | Waste Management | 0.6857 | 1-5 | 1*-3** | 1-6 | 1*-2**-3* |
| NYSE3 | AvalonBay Communities | 0.0738* | 1-5 | 1* | 1-5 | ns |
| NYSE3 | Bank of NY Mellon | < 0.0001*** | 1-6 | ns | 1-5 | 5** |
| NYSE3 | Boston Properties | 0.0045*** | 1-5 | 2*-3** | 1-6 | ns |
| NYSE3 | Charles Schwab | 0.0016*** | 1-4 | 5* | 1-5 | 3** |

**Table 3.** Continued

|  | | Runs above and below the mean | Runs up | | Runs down | |
|---|---|---|---|---|---|---|
|  | Stock | Asymptotic p-value | Lenght | Significant | Lenght | Significant |
| NYSE3 | Franklin Resources | 0.0021*** | 1-6 | 2**-3*** | 1-5 | 1**-2* |
| NYSE3 | Loews | < 0.0001*** | 1-5 | 1**-3**-4* | 1-5 | ns |
| NYSE3 | Moody's | 0.0059*** | 1-6 | ns | 1-2-3-4-5-7 | 7*** |
| NYSE3 | Morgan Stanley | 0.1254 | 1-2-3-4-5-7 | 7*** | 1-7 | 7*** |
| NYSE3 | Prudential Financial | 0.0156** | 1-5 | ns | 1-6 | 6** |
| NYSE3 | State Street | 0.0001*** | 1-5 | 3** | 1-5 | ns |
| NYSE3 | U.S. Bancorp | < 0.0001*** | 1-6 | 4* | 1-5 | ns |
| NYSE3 | Western Union | 0.0011*** | 1-5 | 1** | 1-5 | ns |
| NYSE3 | Weyerhaeuser | 0.2587 | 1-5 | ns | 1-2-3-4-5-7 | 7*** |
| NYSE4 | Aetna | 0.0495** | 1-5 | ns | 1-2-3-4-5-7 | 7*** |
| NYSE4 | Anthem | 0.0712* | 1-6 | 5* | 1-2-3-4-5-7 | 7*** |
| NYSE4 | Baxter International | 0.0710* | 1-4 | 2**-5* | 1-6 | ns |
| NYSE4 | Eli Lilly | 0.1002 | 1-5 | ns | 1-5 | ns |
| NYSE4 | Johnson&Johnson | 0.0527* | 1-5 | 4* | 1-5 | ns |
| NYSE4 | Waters | 0.0537* | 1-6 | 3** | 1-5 | ns |
| NYSE4 | Zimmer Biomet Holdings | 0.1948 | 1-5 | 3* | 1-5 | 5*** |
| NYSE5 | HP | 0.6275 | 1-5 | ns | 1-6 | 6** |
| NYSE5 | Juniper Networks | 0.0754* | 1-5 | 3*** | 1-5 | 2**-3**-4* |
| NYSE5 | Oracle | 0.0030*** | 1-5 | 1**-2** | 1-5 | ns |
| NASD1 | Costco Wholesale | 0.9766 | 1-2-3-4-5-7 | 7*** | 1-6 | 1*-4** |
| NASD1 | Hasbro | 0.8950 | 1-4 | 3*-5* | 1-2-3-4-5-7 | 3**-4*-7** |
| NASD1 | PepsiCo | 0.5468 | 1-5 | ns | 1-5 | ns |
| NASD2 | Cintas | 0.0033*** | 1-5 | ns | 1-5 | ns |
| NASD2 | Paccar | 0.0088*** | 1-5 | ns | 1-5 | ns |
| NASD3 | Huntington Bancshares | < 0.0001*** | 1-5 | ns | 1-5 | 5** |
| NASD3 | Zions | 0.4116 | 1-5 | 4** | 1-5 | ns |
| NASD4 | Biogen | 0.0009*** | 1-6 | ns | 1-5 | ns |
| NASD4 | Mylan | 0.7397 | 1-6 | ns | 1-5 | ns |
| NASD5 | Akamai Technologies | 0.1781 | 1-5 | 1**-2* | 1-5 | ns |
| NASD5 | Analog Devices | 0.0189** | 1-5 | ns | 1-6 | 2** |
| NASD5 | Apple | 0.4959 | 1-5 | 3* | 1-6 | 3**-5*** |
| NASD5 | Citrix System | 0.0360** | 1-5 | 1* | 1-6 | 4** |
| NASD5 | Fiserv | < 0.0001*** | 1-5 | 1**-3** | 1-5 | 3*-4* |
| NASD5 | Microsoft | 0.0046*** | 1-4 | 5* | 1-5 | 4** |
| NASD5 | Symantec | 0.0100** | 1-5 | ns | 1-5 | ns |
|  | **S&P 500** | 0.0010*** | 1-5 | 3*-4* | 1-2-3-4-5-7 | 3**-7*** |

Note: * stands for significance at 10%, ** stands for significance at 5%, *** stands for significance at 1%.

The variance ratio test seems to confirm the presence of severe inefficiency in the selected sample and in the S&P 500 index series (Table 4). We select the periods proposed originally by Lo and MacKinley (1988). The null hypothesis of a martingale is strongly rejected, for both individual tests and joint tests, for Clorox, TJX, Walmart, Yum! Brands, Exxon Mobil, AvalonBay Communities, Bank of New York Mellon, Boston Properties, Charles Schwab, Franklin Resources, Loews, Oracle, Hasbro, Biogen, Analog Devices, Citrix System, Microsoft, Symantec and for the S&P 500 index series. For these stocks the rejection of the individual tests it is at 5% for almost every period and for the joint test. These results are in line with the findings of other studies that implemented the same methodology (Lo and MacKinley, 1988; Chang and Ting, 2000). However, it should be noted that Lo and MacKinley used weekly data to minimize potential biases of daily observations associated with nontrading, the bid-ask spread, asynchronous prices, etc. Even if this is true, in our case the choice of daily observations is appropriate because the sampling theory of the test is based wholly on asymptotic approximations, therefore a large number of observations is needed. Furthermore, we can confront these results with other tests to see whether they converge on the same conclusion or not.

Lo and MacKinley (1988) concluded that the evidence given by the variance ratio test shows that weekly stock returns are incompatible with the random walk model; our findings show that also daily stock returns are incompatible with the model. However, the rejections of the hypothesis do not offer any explicit suggestion toward a more plausible model for the data. Lo and MacKinley suggested possible alternative models: for example, a popular hypothesis is that the stock returns process may be described by the sum of a random walk and a stationary mean-reverting component, as in Summers (1986) and in Fama and French (1988), or only by a mean-reverting process (the Ornstein-Uhlenbeck process) as in Shiller and Perron (1985). The authors clarify that the rejection of the null hypothesis does "*not necessarily imply that the stock market is inefficient or that prices are not rational assessments of "fundamental" values. [...] without a more explicit economic model of the price-generating mechanism, a rejection of the random walk hypothesis has few implications for the efficiency of market price formation*" (Lo and MacKinley, 1988, p. 42). However, the test may be interpreted as a rejection of some economic model of price efficiency, and our approach has the advantage of comparing these results with those derived from different methodologies.

**Table 4.** Variance ratio tests (with unbiased variances) of the random walk hypothesis for the sample. The variance ratios of returns are reported with the heteroscedasticity-robust test statistics $z^*(q)$ given in parentheses. Under the random walk null hypothesis, the value of the variance ratio is 1 and the test statistics have a standard normal distribution (asymptotically). For the joint tests the max |z| statistics is reported and in parentheses the period for which it assumes its maximum is indicated; the p-value for the joint tests is computed via approximation using Studentized Maximum Modulus with parameter 4 and infinite degree of freedom.

|   |   | Individual tests | | | | |
|---|---|---|---|---|---|---|
|   | Stock | Period 2 | Period 4 | Period 8 | Period 16 | Joint test |
| NYSE1 | Best Buy | 0.998 (-0.08) | 0.975 (-0.49) | 0.968 (-0.43) | 1.008 (0.08) | 0.49 (4) |
| NYSE1 | Campbell Soup | 0.985 (-0.52) | 0.932 (-1.31) | 0.925 (-0.95) | 0.957 (-0.37) | 1.31 (4) |
| NYSE1 | Carnival | 0.974 (-0.90) | 0.881 (-2.15)** | 0.840 (-1.77)* | 0.811 (-1.42) | 2.15 (4) |
| NYSE1 | Clorox | 0.907 (-2.78)*** | 0.837 (-2.75)*** | 0.770 (-2.63)*** | 0.744 (-2.10)** | 2.79 (2)** |
| NYSE1 | Constellation Brands A | 0.987 (-0.49) | 0.963 (-0.75) | 0.935 (-0.82) | 0.947 (-0.43) | 0.82 (8) |
| NYSE1 | General Mills | 0.967 (-1.30) | 0.888 (-2.20)** | 0.795 (-2.52)** | 0.733 (-2.29)** | 2.52 (8)** |
| NYSE1 | Kroger | 0.972 (-0.71) | 0.942 (-0.92) | 0.907 (-1.09) | 0.867 (-1.19) | 1.19 (16) |
| NYSE1 | Newell Brands | 0.989 (-0.33) | 0.976 (-0.40) | 0.984 (-0.18) | 0.974 (-0.19) | 0.40 (4) |
| NYSE1 | Nordstrom | 1.039 (1.28) | 1.002 (0.03) | 0.950 (-0.51) | 0.967 (-0.23) | 1.28 (2) |
| NYSE1 | Sysco | 0.927** (-2.10) | 0.874 (-2.05)** | 0.850 (-1.61) | 0.829 (-1.27) | 2.10 (2) |
| NYSE1 | TJX | 0.923 (-2.93)*** | 0.832 (-3.21)*** | 0.683 (-3.68)*** | 0.631 (-2.84)*** | 3.68 (8)*** |
| NYSE1 | Walmart | 0.921 (-2.46)** | 0.826 (-2.98)*** | 0.726 (-3.12)*** | 0.729 (-2.12)** | 3.12 (8)*** |
| NYSE1 | Walt Disney | 0.945 (-1.63) | 0.878 (-1.83)* | 0.825 (-1.63) | 0.800 (-1.24) | 1.83 (4) |
| NYSE1 | Yum! Brands | 0.953 (-1.78)* | 0.836 (-3.26)*** | 0.745 (-3.17)*** | 0.698 (-2.50)** | 3.26 (4)*** |
| NYSE2 | MMM | 0.934 (-2.24)** | 0.870 (-2.22)** | 0.822 (-1.88)* | 0.779 (-1.58) | 2.24 (2)* |
| NYSE2 | AES | 0.901 (-2.32)** | 0.805 (-2.46)** | 0.740 (-2.12)** | 0.683 (-1.76)* | 2.46 (4)* |
| NYSE2 | American Electric Power | 0.915 (-2.12)** | 0.830 (-2.21)** | 0.798 (-1.69)* | 0.727 (-1.57) | 2.21 (4) |
| NYSE2 | Cummins | 0.959 (-1.19) | 0.897 (-1.55) | 0.837 (-1.54) | 0.826 (-1.11) | 1.55 (4) |
| NYSE2 | Eaton | 0.992 (-0.31) | 0.986 (-0.29) | 0.979 (-0.25) | 1.006 (0.05) | 0.31 (2) |
| NYSE2 | Exelon | 0.926 (-1.97)** | 0.865 (-1.86)* | 0.812 (-1.64) | 0.736 (-1.52) | 1.97 (2) |
| NYSE2 | Exxon Mobil | 0.869 (-2.54)** | 0.727 (-2.67)*** | 0.640 (-2,23)** | 0.563 (-1.91)* | 2.67 (4)** |
| NYSE2 | FirstEnergy | 0.918 (-2.13)** | 0.854 (-1.97)** | 0.871 (-1.13) | 0.857 (-0.86) | 2.13 (2) |
| NYSE2 | General Dynamics | 0.948 (-1.77)* | 0.967 (-0.583) | 1.032 (0.36) | 1.112 (0.83) | 1.77 (2) |
| NYSE2 | Hess | 0.970 (-0.87) | 0.907 (-1.37) | 0.842 (-1.43) | 0.809 (-1.15) | 1.43 (8) |
| NYSE2 | L3 Technologies | 0.947 (-1.59) | 0.964 (-0.59) | 0.988 (-0.13) | 1.003 (0.02) | 1.59 (2) |
| NYSE2 | Nisource | 0.958 (-1.25) | 0.904 (-1.52) | 0.810 (-1.89)* | 0.731 (-1.82)* | 1.89 (8) |
| NYSE2 | Noble Energy | 0.957 (-1.06) | 0.898 (-1.35) | 0.835 (-1.39) | 0.746 (-1.47) | 1.47 (16) |
| NYSE2 | Occidental Petroleum | 0.922 (-1.96)** | 0.833 (-2.13)** | 0.759 (-1.88)* | 0.664 (-1.74)* | 2.13 (4) |
| NYSE2 | Parker-Hannifin | 0.975 (-0.83) | 0.945 (-0.96) | 0.908 (-1.01) | 0.906 (-0.71) | 1.01 (8) |
| NYSE2 | Pinnacle West Capital | 0.965 (-1.01) | 0.924 (-1.18) | 0.913 (-0.85) | 0.915 (-0.56) | 1.18 (4) |
| NYSE2 | Schlumberger | 0.935 (-1.88)* | 0.880 (-1.87)* | 0.838 (-1.57) | 0.796 (-1.31) | 1.88 (2) |
| NYSE2 | Southwest Airlines | 0.913 (-2.41)** | 0.916 (-1.35) | 0.912 (-0.98) | 0.956 (-0.35) | 2.41 (2) |
| NYSE2 | Textron | 0.987 (-0.35) | 1.014 (0.20) | 1.075 (0.63) | 1.111 (0.60) | 0.63 (8) |
| NYSE2 | Vulcan Materials | 1.040 (1.28) | 0.993 (-0.13) | 0.961 (-0.43) | 0.926 (-0.57) | 1.28 (2) |
| NYSE2 | Waste Management | 0.966 (-0.88) | 0.876 (-1.74)* | 0.789 (-1.93)* | 0.756 (-1.51) | 1.93 (8) |
| NYSE3 | AvalonBay Communities | 0.810 (-4.11)*** | 0.677 (-3.85)*** | 0.566 (-3.25)*** | 0.524 (-2.37)** | 4.11 (2)*** |
| NYSE3 | Bank of NY Mellon | 0.822 (-3.01)*** | 0.696 (-2.99)*** | 0.561 (-2.89)*** | 0.461 (-2.44)** | 3.01 (2)** |
| NYSE3 | Boston Properties | 0.787 (-4.69)*** | 0.661 (-4.00)*** | 0.552 (-3.28)*** | 0.511 (-2.39)** | 4.69 (2)*** |
| NYSE3 | Charles Schwab | 0.883 (-3.23)*** | 0.796 (-3.13)*** | 0.660 (-3.37)*** | 0.629 (-2.48)** | 3.37 (8)*** |
| NYSE3 | Equifax | 0.983 (-0.48) | 0.964 (-0.53) | 0.932 (-0.63) | 0.885 (-0.75) | 0.75 (16) |
| NYSE3 | Franklin Resources | 0.926 (-2.15)** | 0.835 (-2.52)** | 0.786 (-2.02)** | 0.741 (-1.60) | 2.52 (4)** |
| NYSE3 | Loews | 0.853 (-2.76)*** | 0.767 (-2.29)** | 0.650 (-2.12)** | 0.552 (-1.82)* | 2.76 (2)** |
| NYSE3 | Moody's | 0.993 (-0.19) | 0.960 (-0.57) | 0.881 (-1.11) | 0.826 (-1.11) | 1.11 (8) |

**Table 4.** Continued

|  | Stock | Period 2 | Period 4 | Period 8 | Period 16 | Joint test |
|---|---|---|---|---|---|---|
| NYSE3 | Morgan Stanley | 1.005 (0.07) | 0.896 (-0.74) | 0.691 (-1.52) | 0.589 (-1.55) | 1.55 (16) |
| NYSE3 | Prudential Financial | 0.975 (-0.51) | 0.904 (-0.97) | 0.851 (-0.95) | 0.826 (-0.75) | 0.97 (4) |
| NYSE3 | State Street | 0.883 (-1.97)** | 0.750 (-2.17)** | 0.646 (-2.07)** | 0.533 (-1.94)* | 2.17 (4) |
| NYSE3 | U.S. Bancorp | 0.910 (-1.87)* | 0.874 (-1.35) | 0.718 (-1.91)* | 0.630 (-1.70)* | 1.91 (8) |
| NYSE3 | Western Union | 0.924 (-1.92)* | 0.885 (-1.73)* | 0.834 (-1.60) | 0.824 (-1.15) | 1.92 (2) |
| NYSE3 | Weyerhaeuser | 1.014 (0.525) | 0.973 (-0.52) | 0.898 (-1.32) | 0.854 (-1.24) | 1.32 (8) |
| NYSE4 | Aetna | 0.951 (-1.41) | 0.905 (-1.34) | 0.895 (-0.95) | 0.910 (-0.57) | 1.41 (2) |
| NYSE4 | Anthem | 0.999 (-0.04) | 0.991 (-0.17) | 1.011 (0.13) | 1.011 (0.09) | 0.17 (4) |
| NYSE4 | Baxter International | 1.004 (0.11) | 0.953 (-0.79) | 0.894 (-1.23) | 0.874 (-1.05) | 1.23 (8) |
| NYSE4 | Eli Lilly | 0.909 (-2.43)** | 0.844 (-2.10)** | 0.885 (-0.99) | 0.846 (-0.93) | 2.43 (2)* |
| NYSE4 | Johnson&Johnson | 0.926 (-1.99)** | 0.845 (-2.05)** | 0.819 (-1.54) | 0.789 (-1.27) | 2.05 (4) |
| NYSE4 | Waters | 0.983 (-0.50) | 0.926 (-1.19) | 0.870 (-1.38) | 0.784 (-1.60) | 1.60 (16) |
| NYSE4 | Zimmer Biomet Holdings | 1.017 (0.55) | 1.005 (0.09) | 0.953 (-0.50) | 0.958 (-0.30) | 0.55 (2) |
| NYSE5 | HP | 1.005 (0.16) | 0.963 (-0.58) | 0.882 (-1.17) | 0.845 (-1.05) | 1.17 (8) |
| NYSE5 | Juniper Networks | 0.954 (-1.69)* | 0.903 (-1.97)** | 0.866 (-1.73)* | 0.864 (-1.16) | 1.97 (4) |
| NYSE5 | Oracle | 0.914 (-2.85)*** | 0.862 (-2.42)** | 0.811 (-2.13)** | 0.742 (-1.95)* | 2.85 (2)** |
| NASD1 | Costco Wholesale | 0.966 (-1.12) | 0.914 (-1.48) | 0.830 (-1.84)* | 0.817 (-1.37) | 1.84 (8) |
| NASD1 | Hasbro | 0.942 (-2.23)** | 0.872 (-2.63)*** | 0.822 (-2.34)** | 0.790 (-1.88)* | 2.63 (4)** |
| NASD1 | PepsiCo | 0.953 (-1.01) | 0.905 (-1.17) | 0.853 (-1.21) | 0.830 (-1.01) | 1.21 (8) |
| NASD2 | Cintas | 0.953 (-1.66)* | 0.932 (-1.25) | 0.880 (-1.37) | 0.838 (-1.24) | 1.66 (2) |
| NASD2 | Paccar | 0.953 (-1.56) | 0.887 (-1.98)** | 0.809 (-2.02)** | 0.778 (-1.52) | 2.02 (8) |
| NASD3 | Huntington Bancshares | 0.927 (-1.20) | 0.856 (-1.28) | 0.717 (-1.65)* | 0.741 (-1.04) | 1.65 (8) |
| NASD3 | Zions | 0.985 (-0.36) | 0.912 (-1.09) | 0.842 (-1.25) | 0.774 (-1.21) | 1.25 (8) |
| NASD4 | Biogen | 0.961 (-1.73)* | 0.891 (-2.57)** | 0.880 (-1.86)* | 0.807 (-2.03)** | 2.57 (4)** |
| NASD4 | Mylan | 1.042 (1.44) | 1.059 (0.95) | 0.999 (-0.01) | 0.931 (-0.48) | 1.44 (2) |
| NASD5 | Akamai Technologies | 0.981 (-0.79) | 0.949 (-1.14) | 0.892 (-1.54) | 0.851 (-1.44) | 1.54 (8) |
| NASD5 | Analog Devices | 0.933 (-2.54)** | 0.863 (-2.69)*** | 0.817 (-2.32)** | 0.749 (-2.15)** | 2.69 (4)** |
| NASD5 | Apple | 0.999 (-0.03) | 0.979 (-0.36) | 1.016 (0.18) | 1.042 (0.30) | 0.36 (4) |
| NASD5 | Citrix System | 0.913 (-3.30)*** | 0.853 (-2.95)*** | 0.821 (-2.25)** | 0.761 (-2.02)** | 3.30 (2)*** |
| NASD5 | Fiserv | 0.923 (-1.88)* | 0.885 (-1.57) | 0.817 (-1.63) | 0.781 (-1.33) | 1.88 (2) |
| NASD5 | Microsoft | 0.930 (-2.13)** | 0.851 (-2.48)** | 0.737 (-2.82)*** | 0.666 (-2.43)** | 2.82 (8)** |
| NASD5 | Symantec | 0.912 (-2.28)** | 0.808 (-2.87)*** | 0.774 (-2.43)** | 0.780 (-1.78)* | 2.87 (4)** |
|  | **S&P 500** | 0.902 (-2.59)*** | 0.809 (-2.48)** | 0.733 (-2.14)** | 0.690 (-1.65)* | 2.59 (2)** |

*Note: * stands for significance at 10%, ** stands for significance at 5%, *** stands for significance at 1%.*

Finally, we can compare all the results to conclude whether a stock is efficient or not. At this point, unfortunately, a certain degree of subjectivity must be introduced, however, the null hypothesis of informational efficiency of a stock will be rejected only when the evidence is strongly against it. Table 5 synthesizes the results of the various statistical tests performed, to assess whether the selected stocks follow a random walk. Runs tests and variance ratio tests give us an answer about the randomness of financial returns series. The answer "No" means that the series is non-random and therefore the random walk hypothesis should be rejected. Given the results of our simulations, we considered inefficient, stocks for which the runs above and below the mean was significant at 1%, or at 5% but with some runs up or down sequence significant. The only exception to this rule is Juniper Networks: in this case the runs above and below the mean test is significant only at 10% but that are 4 significant sequence of runs up and down, three of which significant at 5%. For the variance ratio individual tests, we rejected the null hypothesis when one or more individual tests were significant at 5%. The global result of these tests is given by the aggregation of runs tests and variance ratio tests (individual and joint): we conclude that a stock is inefficient only if all the three tests reject the null hypothesis of a random walk series (a very conservative, yet prudent, choice).

From the last column of Table 5 it results that the 18.6% of the selected stocks traded on the NYSE is inefficient, whereas the 25% of the selected stocks traded on the NASDAQ is inefficient. The fact that the NASDAQ is a computer-based market with orders processed electronically and that stocks on the NASDAQ are considered to be more volatile and growth oriented could explain the higher level of inefficiency. The most inefficient stratus of the sample was the financial and real estate sector of stocks traded on the NYSE, with the 35.7% of stocks inefficient (5 stocks over 14). Once again, the explanation may be linked with the volatility of this sector and to the great amount of information needed to successfully invest.

Table 6 shows the computed confidence intervals for proportion. The higher level of inefficiency is given by the individual variance ratio tests: the proportion, with a confidence of 95%, of inefficient stocks continuously traded as S&P 500 Index components from Jan 2, 2008 to July 24, 2018 is estimated to be between the 34.23% and the 53.77%. The range given using runs tests is also remarkable, whereas the variance ratio joint test is the most conservative methodology. Once we gathered all the results, we estimated a confidence interval for the proportion of inefficient stocks with a lower limit of 12.13% and an upper limit of 27.87%. This range is calculated considering inefficient stocks as signalled by both the runs test and the variance ratio test, therefore the empirical evidence supporting this result is very strong. This proofs that a certain degree of informational inefficiency exists in the U.S. stock market.

**Table 5.** Summary of the statistical tests. The columns report a "Yes" if the financial returns series is informationally efficient (it follows a random walk), a "No" otherwise. The column "Global result" is given by the aggregation of runs tests and variance ratio tests: stocks are considered inefficient only if the performed tests converge on the same conclusion.

|  | | | Variance Ratio | | |
|---|---|---|---|---|---|
|  | Stock | Runs test | Individual Tests | Joint Test | Global result |
| NYSE1 | Best Buy | Yes | Yes | Yes | Yes |
| NYSE1 | Campbell Soup | No | Yes | Yes | Yes |
| NYSE1 | Carnival | Yes | No | Yes | Yes |
| NYSE1 | Clorox | No | No | No | No |
| NYSE1 | Constellation Brands A | Yes | Yes | Yes | Yes |
| NYSE1 | General Mills | No | No | No | No |
| NYSE1 | Kroger | Yes | Yes | Yes | Yes |
| NYSE1 | Newell Brands | Yes | Yes | Yes | Yes |
| NYSE1 | Nordstrom | Yes | Yes | Yes | Yes |
| NYSE1 | Sysco | Yes | No | Yes | Yes |
| NYSE1 | TJX | No | No | No | No |
| NYSE1 | Walmart | Yes | No | No | Yes |
| NYSE1 | Walt Disney | Yes | Yes | Yes | Yes |
| NYSE1 | Yum! Brands | No | No | No | No |
| NYSE2 | MMM | Yes | No | Yes | Yes |
| NYSE2 | AES | No | No | Yes | Yes |
| NYSE2 | American Electric Power | Yes | No | Yes | Yes |
| NYSE2 | Cummins | Yes | Yes | Yes | Yes |
| NYSE2 | Eaton | Yes | Yes | Yes | Yes |
| NYSE2 | Exelon | Yes | No | Yes | Yes |
| NYSE2 | Exxon Mobil | No | No | No | No |
| NYSE2 | FirstEnergy | Yes | No | Yes | Yes |
| NYSE2 | General Dynamics | Yes | Yes | Yes | Yes |
| NYSE2 | Hess | Yes | Yes | Yes | Yes |
| NYSE2 | L3 Technologies | Yes | Yes | Yes | Yes |
| NYSE2 | Nisource | Yes | Yes | Yes | Yes |
| NYSE2 | Noble Energy | No | Yes | Yes | Yes |
| NYSE2 | Occidental Petroleum | Yes | No | Yes | Yes |
| NYSE2 | Parker-Hannifin | Yes | Yes | Yes | Yes |
| NYSE2 | Pinnacle West Capital | Yes | Yes | Yes | Yes |
| NYSE2 | Schlumberger | Yes | Yes | Yes | Yes |
| NYSE2 | Southwest Airlines | No | No | Yes | Yes |
| NYSE2 | Textron | Yes | Yes | Yes | Yes |
| NYSE2 | Vulcan Materials | Yes | Yes | Yes | Yes |
| NYSE2 | Waste Management | Yes | Yes | Yes | Yes |
| NYSE3 | AvalonBay Communities | Yes | No | No | Yes |
| NYSE3 | Bank of NY Mellon | No | No | No | No |
| NYSE3 | Boston Properties | No | No | No | No |
| NYSE3 | Charles Schwab | No | No | No | No |
| NYSE3 | Equifax | Yes | Yes | Yes | Yes |
| NYSE3 | Franklin Resources | No | No | No | No |
| NYSE3 | Loews | No | No | No | No |
| NYSE3 | Moody's | No | Yes | Yes | Yes |

**Table 5.** Continued

|  | Stock | Runs test (returns) | Variance Ratio Individual Tests (returns) | Joint Test (returns) | Global result (returns) |
|---|---|---|---|---|---|
| NYSE3 | Morgan Stanley | Yes | Yes | Yes | Yes |
| NYSE3 | Prudential Financial | No | Yes | Yes | Yes |
| NYSE3 | State Street | No | No | Yes | Yes |
| NYSE3 | U.S. Bancorp | No | Yes | Yes | Yes |
| NYSE3 | Western Union | No | Yes | Yes | Yes |
| NYSE3 | Weyerhaeuser | Yes | Yes | Yes | Yes |
| NYSE4 | Aetna | No | Yes | Yes | Yes |
| NYSE4 | Anthem | Yes | Yes | Yes | Yes |
| NYSE4 | Baxter International | Yes | Yes | Yes | Yes |
| NYSE4 | Eli Lilly | Yes | No | Yes | Yes |
| NYSE4 | Johnson&Johnson | Yes | No | Yes | Yes |
| NYSE4 | Waters | Yes | Yes | Yes | Yes |
| NYSE4 | Zimmer Biomet Holdings | Yes | Yes | Yes | Yes |
| NYSE5 | HP | Yes | Yes | Yes | Yes |
| NYSE5 | Juniper Networks | No | No | Yes | Yes |
| NYSE5 | Oracle | No | No | No | No |
| NASD1 | Costco Wholesale | Yes | Yes | Yes | Yes |
| NASD1 | Hasbro | Yes | No | No | Yes |
| NASD1 | PepsiCo | Yes | Yes | Yes | Yes |
| NASD2 | Cintas | No | Yes | Yes | Yes |
| NASD2 | Paccar | No | No | Yes | Yes |
| NASD3 | Huntington Bancshares | No | Yes | Yes | Yes |
| NASD3 | Zions | Yes | Yes | Yes | Yes |
| NASD4 | Biogen | No | No | No | No |
| NASD4 | Mylan | Yes | Yes | Yes | Yes |
| NASD5 | Akamai Technologies | Yes | Yes | Yes | Yes |
| NASD5 | Analog Devices | No | No | No | No |
| NASD5 | Apple | Yes | Yes | Yes | Yes |
| NASD5 | Citrix System | No | No | No | No |
| NASD5 | Fiserv | No | Yes | Yes | Yes |
| NASD5 | Microsoft | No | No | No | No |
| NASD5 | Symantec | Yes | No | No | Yes |
|  | **S&P 500** | No | No | No | No |
|  | **Tot. Inefficient Stocks** | 30 | 33 | 19 | 15 |

**Table 6.** Confidence intervals for the proportion (with finite population correction) of inefficient S&P 500 stocks in the period 2008-2018. The confidence level is 95%.

| Confidence interval using Runs tests | Confidence interval using Individual Variance Ratio tests |
|---|---|
| [30.36%; 49.64%] | [34.23%; 53.77%] |
| Confidence interval using Joint Variance Ratio Tests | Confidence interval using all the tests |
| [16.77%; 33.89%] | [12.13%; 27.87%] |

## 4. Conclusions

In this work we tried to further validate the claim of inefficiencies in financial markets investigating the degree of informational efficiency combining classical techniques with a novel methodology. We selected a sample of stocks continuously traded as components of the S&P 500 Index from January 2, 2008 to July 24, 2018 and via the use of runs tests and variance ratio tests we estimated the proportion of inefficient stocks in the index to be between the 12.13% and the 27.87%. These findings are in line with the works of many authors (for example, see: LeRoy and Porter, 1981; De Bondt and Thaler, 1987; Lo and MacKinlay, 1988) and provide new evidence regarding the existence of a certain degree of inefficiency in financial markets. This evidence should not be interpreted as an invite to challenge arrogantly the market, but rather as further proof that markets are made by humans, therefore they are not flawless. Of course, if you have to decide to who you should entrust your savings, we think that an analyst will make a better job than a blindfolded monkey. In the eternal fight between analysts and markets, however, we cannot still declare a winner: even though market has won most of the battles, the war is still open.

**Funding**

This research did not receive any specific grant from funding agencies in the public, commercial, or not-for-profit sectors.

# Supplementary Material for the paper "Blindfolded monkeys or financial analysts: who is worth your money?"

# Appendix A

We show how to derive Yamane sample size formula (4) in the paper from equation (2) using Tejada and Punzalan (2012)[8] approach and Cochrane (1953) notation. Remembering that

$$n_0 = \frac{t^2 pq}{d^2} \qquad (2)$$

$$n = \frac{n_0}{1 + \frac{n_0}{N}} \qquad (3)$$

we assume a 95% degree of confidence so that *t* is approximately equal to 2. Because we have not any prior knowledge about *P* (the quantity which we would like to measure), the conservative approach suggests maximizing

$$p(1-p) = \frac{1}{4} - \left(\frac{1}{2} - p\right)^2$$

that assumes its maximum when *p* = 0.5, so that the subtrahend is equal to zero.
Substituting this value in (2)

$$n_0 = \frac{2^2 0.5(1 - 0.5)}{d^2} = \frac{1}{d^2} \qquad (2a)$$

substituting (2a) in (3) gives

$$n = \frac{\frac{1}{d^2}}{1 + \frac{\frac{1}{d^2}}{N}} = \frac{N}{1 + Nd^2} \qquad (4)$$

this completes the proof.

---

[8] Tejada, J. J. and Punzalan, J. R. B. (2012). "On the Misuse of Slovin's Formula", *The Philippine Statistician*, 61 (1), pp. 129-136.